\documentclass[aps,preprint,epsfig,rotate]{revtex4}
\usepackage{graphicx}
\usepackage{bm}
\usepackage{epsfig}

\input epsf
\begin{document}

\title{Accurate computations of bound state properties in three- and four-electron atomic systems in the basis of multi-dimensional gaussoids.} 

 \author{Alexei M. Frolov}
 \email[E--mail address: ]{afrolov@uwo.ca}

 \affiliation{Department of Applied Mathematics, University of Western Ontario,
              London, Ontario, N6H 5E5, Canada}

\author{David M. Wardlaw}
 \email[E--mail address: ]{dwardlaw@mun.ca}

\affiliation{Department of Chemistry, Memorial University of Newfoundland, St.John's, 
             Newfoundland and Labrador, A1C 5S7, Canada}

\date{\today}

\begin{abstract}

Results of accurate computations of bound states in three- and four-electron atomic systems are discussed. Bound state properties of the four-electron lithium ion Li$^{-}$ 
in its ground $2^{2}S-$state are determined from the results of accurate, variational computations. We also consider a closely related problem of accurate numerical 
evaluation of the half-life of the beryllium-7 isotope. This problem is of paramount importance for modern radiochemistry.


\end{abstract}

\maketitle

\section{Introduction}

In this communication we consider the bound states properties of the negatively charged Li$^{-}$ ion in its ground $2^1S(L = 0)-$state, or $2^1S-$state, for short. It is well 
known that the $2^{1}S-$state is the only bound state in this ion. The electronic structure of this state in Li$^{-}$ corresponds to the $1s^2 2s^2$ electron configuration. The 
negatively charged lithium ion has become of interest in numerous applications, since formation of these ions is an important step for workability of lithium and/or lithium-ion 
electric batteries (see, e.g., \cite{Ref1}, \cite{Ref2} and references therein). Both lithium and lithium-ion batteries are very compact, relatively cheap and reliable sources 
of constant electric current which are widely used in our everyday life and in many branches of modern industry. However, it appears that the Li$^{-}$ ion is not a well studied 
atomic system. Indeed, many bound state properties of this ion have not been evaluated at all even for an isolated Li$^{-}$ ion in vacuum. In reality, it is crucial to know its 
bound state properties in the presence of different organic acids which are extensively used in lithium-ion batteries. 

Our goal in this study is to determine various bound state properties of the four-electron (or five-body) Li$^{-}$ ion and compare them with the corresponding properties of the 
neutral Li atom in its ground $2^{2}S-$state. It should be mentioned that many of the bound state properties of the Li$^{-}$ ion have not been evaluated in earlier studies. The 
negatively charged Li$^{-}$ ion is here described by the non-relativistic Schr\"{o}dinger equation $H \Psi = E \Psi$, where $H$ is the non-relativistic Hamiltonian, $E(< 0)$ is the 
eigenvalue and $\Psi$ is the bound state wave function of the Li$^{-}$ ion. Without loss of generality we shall assume that the bound state wave function $\Psi$ has the unit norm. 
The non-relativistic Hamiltonian $H$ of an arbitrary four-electron atom/ion takes the form (see, e.g., \cite{LLQ})
\begin{eqnarray}
 H = -\frac{\hbar^2}{2 m_e} \Bigl[\nabla^2_1 + \nabla^2_2 + \nabla^2_3 + \nabla^2_4 + \frac{m_e}{M} \nabla^2_5 \Bigr] - Q e^2 \sum^{4}_{i=1} \frac{1}{r_{15}} 
 + e^{2} \sum^{3}_{i=1} \sum^{4}_{j=2 (j>i)} \frac{1}{r_{ij}} \; \; \; , \; \; \label{Hamil}
\end{eqnarray}
where $\hbar = \frac{h}{2 \pi}$ is the reduced Planck constant, $m_e$ is the electron mass and $e$ is the electric charge of an electron. In this equation and everywhere below in 
this study the subscripts 1, 2, 3, 4 designate the four atomic electrons $e^-$, while the subscript 5 (= $N$) denotes the heavy atomic nucleus with the mass $M$ ($M \gg m_e$), and 
the positive electric (nuclear) charge is $Q e$. The notation $r_{ij} = \mid {\bf r}_i - {\bf r}_j \mid = r_{ji}$ in Eq.(\ref{Hamil}) and everywhere below stands for the interparticle 
distances between particles $i$ and $j$. These distances are also called the relative coordinates to differentiate them from the three-dimensional coordinates ${\bf r}_i$, 
which are the Cartesian coordinates of the particle $i$. In Eq.(\ref{Hamil}) and everywhere below in this work we shall assume that $(ij)$ = $(ji)$ = (12), (13), (14), (15), (23), 
(24), (25), (34), (35) and (45), for four-electron atomic systems and particle 5 means the atomic nucleus. Analogously, for three-electron atomic systems we have $(ij)$ = $(ji)$ = 
(12), (13), (14), (23), (24) and (34), where particle 4 is the atomic nucleus. Below only atomic units $\hbar = 1, \mid e \mid = 1, m_e = 1$ are employed. In these units the 
explicit form of the Hamiltonian $H$, Eq.(\ref{Hamil}), is simplified and takes the form
\begin{eqnarray}
 H = -\frac12 \Bigl[\nabla^2_{1} + \nabla^2_{2} + \nabla^2_{3} + \nabla^2_{4} + \frac{m_e}{M} \nabla^2_{5} \Bigr] - Q \sum^{4}_{i=1} \frac{1}{r_{i5}} + \sum^{3}_{i=1} 
 \sum^{4}_{j=2 (j>i)} \frac{1}{r_{ij}} \; \; \; , \; \; \label{Hamil1}
\end{eqnarray}
where $Q$ is the nuclear charge of the central positively charged nucleus. For the negatively charged Li$^{-}$ ion we have $Q = 3$. Note that the stability of the bound $2^1S-$state 
in the Li$^{-}$ ion means stability against its dissociation (or ionization) Li$^{-} \rightarrow$ Li$(2^2S)$ + $e^{-}$, where the notation Li$(2^2S)$ means the lithium atom in its ground 
(doublet) $2^2S-$state. In general, the bound state properties of the neutral Li atom in its ground $2^{2}S-$state are important to predict and approximately evaluate analogous
bound state properties of the negatively charged Li$^{-}$ ion. In this study such evaluations are considered carefuly, but first of all we need to describe our method which is used
to construct accurate wave functions for four-electron atomic systems. This problem is considered in the next Section.   

\section{Variational wave functions for four- and three-electron atomic systems}

To determine accurate solutions of the non-relativistic Schr\"{o}dinger equation $H \Psi = E \Psi$ in this study we apply variational expansion written in multi-dimensional gaussoids. 
Each of these gaussoids explicitly depends upon a number of the relative coordinates $r_{ij}$. For four-electron atoms and ions there are ten relative coordinates: $r_{ij} = r_{12}, 
r_{13}, r_{14}, r_{15}, r_{23}, \ldots, r_{45}$. In particular, for the singlet ${}^{1}S(L = 0)-$states in four-electron atomic systems the variational expansion in multi-dimensional 
gaussoids takes the form (see, e.g., \cite{KT}, \cite{Fro1}):
\begin{eqnarray}
  \psi(L = 0; S = 0) = \sum^{N_A}_{i=1} C_i {\cal A}_{1234} [\exp(-\alpha_{ij} r^2_{ij}) \chi^{(1)}_{S=0}] + 
  \sum^{N_B}_{i=1} G_i {\cal A}_{1234} [\exp(-\beta_{ij} r^2_{ij}) \chi^{(2)}_{S=0}] \; \; \; \label{equa1}
\end{eqnarray}
where  ${\cal A}_{1234}$ is the complete four-electron anti-symmetrizer, $C_i$ (and $G_i$) are the linear variational coefficients of the variational function, while $\alpha_{ij}$,  
where (ij) = (12), (13), $\ldots$, (45), are the ten non-linear parameters in the radial function associated with the $\chi^{(1)}_{S=0}$ spin function. Analogously, the notation 
$\beta_{ij}$ stands for other ten non-linear parameters in the radial function associated with the $\chi^{(2)}_{S=0}$ spin function. Note that these two sets of non-linear parameters
$\alpha_{ij}$ and $\beta_{ij}$ must be varied independently in calculations. Notations $\chi^{(1)}_{S=0}$ and $\chi^{(2)}_{S=0}$ in Eq.(\ref{equa1}) designate the two independent spin 
functions which can be considered for the singlet $2^1S-$state, or $(2^{1}S \mid 1s^2 2s^2)-$electron configuration. The explicit forms of these two spin functions are:
\begin{eqnarray}
  \chi^{(1)}_{S=0} &=& \alpha \beta \alpha \beta + \beta \alpha \beta \alpha - \beta \alpha \alpha \beta - \alpha \beta \beta \alpha \;\;\; \label{spin1} \\
 \chi^{(2)}_{S=0} &=& 2 \alpha \alpha \beta \beta + 2 \beta \beta \alpha \alpha - \beta \alpha \alpha \beta - \alpha \beta \beta \alpha - \beta \alpha \beta \alpha 
  - \alpha \beta \alpha \beta \;\;\; \label{spin2} 
\end{eqnarray}
where $\alpha$ and $\beta$ are the single-electron spin-up and spin-down functions \cite{Dirac}. In numerical calculations of the total energies and other spin-independent properties 
(i.e. expectation values) one can always use just one spin function, e.g., $\chi^{(1)}_{S=0}$ from Eq.(\ref{spin1}). It follows from the fact that the Hamiltonian Eq.(\ref{Hamil1}) 
does not depend explicitly upon the electron spin and/or any of its components. 

For three-electron atomic systems considered in this study, e.g., for the Li-atom, the analogous expansion in multi-dimensional gaussoids is written in the form \cite{KT}, \cite{Fro1}
\begin{eqnarray}
  \psi(L = 0; S = \frac12) = \sum^{N_A}_{i=1} C_i {\cal A}_{123} [\exp(-\alpha_{ij} r^2_{ij}) \chi^{(1)}_{S=\frac12}] + 
  \sum^{N_B}_{i=1} G_i {\cal A}_{123} [\exp(-\beta_{ij} r^2_{ij}) \chi^{(2)}_{S=\frac12}] \; \; \; \label{equa2}
\end{eqnarray}
where ${\cal A}_{123}$ is the complete three-electron (or three-particle) anti-symmetrizer, $C_i$ (and $G_i$) are the linear variational coefficients of the variational function, while 
$\alpha_{ij}$,  where (ij) = (12), (13), $\ldots$, (34), are the six non-linear parameters for three-electron atomic systems. In these notations the notations/indexes 1, 2, 3 designate
three atomic electrons, while 4 means heavy atomic nucleus. Analogously, the notation $\beta_{ij}$ stands for other six non-linear parameters which must also be varied (independently 
of $\alpha_{ij}$) in calculations. Notations $\chi^{(1)}_{S=\frac12}$ and $\chi^{(2)}_{S=\frac12}$ in Eq.(\ref{equa1}) designate the two independent spin functions which can be 
considered for the doublet $2^{2}S-$state, or $(2^{2}S \mid 1s^2 2s^1)-$electron configuration. The explicit forms of these two spin functions are:
\begin{eqnarray}
  \chi^{(1)}_{S=\frac12} &=& \alpha \beta \alpha - \beta \alpha \alpha \;\;\; \label{spin1a} \\
 \chi^{(2)}_{S=\frac12} &=& 2 \alpha \alpha \beta - \beta \alpha \alpha - \alpha \beta \alpha  \;\;\; \label{spin2a} 
\end{eqnarray}

The Hamiltonian of the three-electron atomic system (e.g., Li-atom) is 
\begin{eqnarray}
 H = -\frac12 \Bigl[\nabla^2_{1} + \nabla^2_{2} + \nabla^2_{3} + \frac{m_e}{M} \nabla^2_{4} \Bigr] - Q \sum^{3}_{i=1} \frac{1}{r_{i4}} + \sum^{2}_{i=1} 
 \sum^{3}_{j=2 (j>i)} \frac{1}{r_{ij}} \; \; \; , \; \; \label{Hamil3}
\end{eqnarray}
where all notations have the same meaning as in Eq.(\ref{Hamil1}). The only difference with Eq.(\ref{Hamil1}) is the fact that here we are dealing with the three-electron atomic systems. 
In particular, the index 4 means the heavy atomic nucleus with the electric charge $Q$ (or $Q e$). Note also that the explicit forms of the three- and four-particle anti-symmetrizers, 
optimization of the non-linear parameters and other important steps in construction of the variational expansions Eqs.(\ref{equa1}) - (\ref{equa2}) for four- and three-electron atoms, 
respectively, have been described in detail in a large number of papers (see, e.g., \cite{Lars} - \cite{FroWa3}, \cite{Ruiz} and references therein). Here we do not want to repeat these 
descriptions of the four- and three-electron variational methods which are used for accurate numerical calculations of various few-electron atoms and ions. In the next two Sections we 
discuss results obtained for the negatively charged four-electron Li$^{-}$ ion and neutral three-electron Li atom, respectively.

\section{Results for the negatively charged lithium ion}
     
As mentioned above in this paper we consider the ground $2^1S-$state of the Li$^{-}$ ion with the infinitely heavy nucleus (i.e., the ${}^{\infty}$Li$^{-}$ ion). Our goal is to determine the 
total energy of this ion and expectation values of some of its properties. Such properties include a few powers of interparticle distances $\langle r^{n}_{ij} \rangle$, where $n = -2, -1, 1, 
2, 3, 4$ (for $n = 0$ each of these expectation values equals unity), electron-nucleus and electron-electron delta-functions, single electron kinetic energy $\langle \frac12 {\bf p}^2_e 
\rangle$, and a few others. As shown in the Appendix the electron-nucleus and electron-electron kinetic correlations $\langle {\bf p}_e \cdot {\bf p}_N \rangle$ and $\langle {\bf p}_e \cdot 
{\bf p}_e \rangle$ are not truly independent atomic properties. Therefore, there is no need to include those expectation values in Table I. Table I also includes the bound state properties 
of the ground $2^{2}S-$state in the neutral Li atom, which is a three-electron atomic system. All these properties are expressed in atomic units. 

The expectation values of the different bound state properties computed for the four-electron Li$^{-}$ ion (or ${}^{\infty}$Li$^{-}$ ion) can be compared with the similar properties of the
ground $2^2S-$state of the three-electron Li atom (or ${}^{\infty}$Li atom). As follows from Table I there are some substantial differences in the electron-nucleus and electron-electron 
distances $\langle r_{eN} \rangle$ and $\langle r_{ee} \rangle$ in the four-electron Li$^{-}$ ion and three-electron Li atom. For the Li$^{-}$ ion these distances are significantly larger 
than for the neutral Li atom. The same conclusion is correct for all positive powers of these inter-particle distances, i.e. for the $\langle r^{k}_{eN} \rangle$ and $\langle r^{k}_{ee} 
\rangle$  expectation values (here $k$ is integer and $k \ge 2$). For the negative powers of interparticle distances, i.e. for the $\langle r^{k}_{eN} \rangle$ and $\langle r^{k}_{ee}  
\rangle$ expectation values (here $k$ is integer and $k \le -1$) the situation is opposite. This is an indication of the known fact that the Li$^{-}$ ion is a weakly-bound, 
four-electron system atomic system. This fact can be confirmed by calculation of the following dimensionless ratio 
\begin{equation}
   \epsilon = \frac{E({\rm Li}^{-}) - E({\rm Li})}{E({\rm Li}^{-})} \approx 0.00301
\end{equation}
where $E($Li$^{-})$ and $E($Li) are the total energies of the negatively charged Li$^{-}$ ion in the ground $2^{1}S-$state and Li atom in the ground $2^2S-$state. A very small value of this 
parameter $\epsilon$, which here is significantly less that 0.01 (or 1 \%), is a strong indication that the Li$^{-}$ ion is an extremely weakly-bound atomic system. This allows 
one to represent the internal structure of the bound state in the Li$^{-}$ ion as a motion of one electron in the `central' field created by the infinitely heavy Li atom. In other words, 
the electronic structure of this ion is $1s^2 2s^2$ and one of the two outer-most electrons moves at very large distances from the central nucleus. In reality, this representation 
is only approximate, since, e.g., there is an exchange symmetry between two electrons in the $2s^2$ shell. Nevertheless, such a `cluster' structure can be useful to predict and 
explain a large number of bound state properties of the Li$^{-}$ ion. For instance, consider the expectation value of the inverse electron-nucleus distance, i.e. $\langle r^{-1}_{eN}
\rangle$. From the definition of this expectation value we write the following expression
\begin{equation}
  \langle r^{-1}_{eN} \rangle = \frac{1}{4} \Bigl( \langle r^{-1}_{1N} \rangle + \langle r^{-1}_{2N} \rangle + \langle r^{-1}_{3N} \rangle + \langle r^{-1}_{4N} \rangle \Bigr)
  \label{eqsym1}
\end{equation}
where all expectation values in the right-hand side are determined without any additional symmetrization between four electrons. As mentioned above the Li$^{-}$ ion has a sharp cluster 
structure and its fourth electron is located on avarage far away from the central nucleus. This means that $\langle r^{-1}_{4N} \rangle \approx 0$. In this case it follows from 
Eq.(\ref{eqsym1}) that 
\begin{equation}
  \langle r^{-1}_{eN} \rangle = \frac{3}{4} \langle r^{-1}_{1N} \rangle = \frac{3}{4} \langle r^{-1}_{eN} \rangle \approx \frac{3}{4} \langle r^{-1}_{eN} \rangle_{{\rm Li}} \label{eqsym2}
\end{equation}
where $\langle r^{-1}_{eN} \rangle_{{\rm Li}}$ is the corresponding expectation value for the neutral Li-atom. It is clear that this equality is only approximate. Analogous approximate 
evaluations can be obtained for some other properties, e.g., for the expectation values of all delta-functions and inverse powers of electron-nucleus and electron-electron distances. 

Table I contains a large number of bound state properties of the negatively charged Li$^{-}$ ion. Numerical values of these properties are of interest in various scientific and technical 
applications, including quite a few applications to electro-chemistry of the lithium and lithium-ion batteries. Our expectation values form a complete set of numerical values which 
can be useful in analysis of different macroscopic systems containing neutral lithium atoms and negatively charged lithium ions.     

\section{Accurate computations of the ground states in heavy three-electron ions}

For three-electron atoms and ions one finds a large number of interesting problems which have not been solved in earlier studies. Here we consider the two following problems: (1) accurate
computations of the ground state ($2^{2}S-$state) energies for some heavy three-electron ions (Sc$^{18+}$ - Ni$^{25+}$), and (2) accurate numerical evaluation of some basic geometrical
properties (expectation values) for these three-electron ions. In these computations we have assumed that all atomic nuclei are infinitely heavy. Results of our computations of these ions 
(ground doublet $2^{2}S-$states) can be found in Table II (in atomic units). It should be mentioned that the overall accuracy of the variational expansion of six-dimensional gaussoids is 
substantially greater than the analogous accuracy achieved with a similar variational expansion for the four-electron atoms/ions. In reality, the accuracy of our procedure has been 
restricted by the double-precision accuracy of our optimization code and results in a maximal accuracy for the total energy of $3 \cdot 10^{-12} - 1 \cdot 10^{-13}$ $a.u$. This maximal 
accuracy was observed in bound state calculations of the heavy ions (all three-electron ions after Cl$^{14+}$). 

Our current results (energies) obtained for heavy three-electron ions allow us to complete the Table (published in \cite{Fro2015}) of the bound states energies of different few-electron 
atomic systems (see Table III). The original Table in \cite{Fro2015} was based on our highly accurate results for two-electron systems and also on the results from \cite{Tamb} and 
\cite{Sims} for three- and four-electron atoms/ions, respectively. In general, the main idea from \cite{Fro2015} works well for few-electron atoms and ions. However, the overall accuracy
of our predictions for total energies of few-electron atomic systems is not very high, since the total energies of the four-electron atoms and ions have been determined \cite{Sims} to the 
accuracy which is substantially lower than the analogous accuracy achieved for two- and three-electron atomic systems. For instance, by using data from the last column of Table III and 
asympotic formulas for $Q^{-1}$ expansion (see, e.g., \cite{Fro2015}) one can obatin only very approximate value of the total energy of the ${}^{\infty}$Li$^{-}$ ion. Furthermore, the 
total energies of some four-electron ions, e.g., in the case of Ar, differ substantially from numerical values known from other papers (see, e.g., \cite{Mal}).   

Another aim of this study was to perform accurate numerical evaluations of bound state properties for a number of heavy three-electron ions. Here we chose the same multi-charged 
three-electron ions Sc$^{18+}$ - Ni$^{25+}$ in their ground doublet $2^{2}S-$states. Results of these calculations can be found in Table II (in atomic units), where a number of electron-electron 
and electron-nucleus $\langle r^{k}_{ij} \rangle$ expectation values (for $k$ = -2, -1 and 1) are shown. As follows from Table II the computed expectation values smoothly vary with the electric 
charge of the atomic nucleus $Q$. In other words, these expectation values are uniform functions of $Q$. Formally, we can propose a number of relatively simple interpolation formulas (upon $Q$)
for these expectation values. 

\section{On the half-life of the beryllium-7 isotope}

Results of our accurate computations of the ground $2^1S-$state in the weakly-bound Li$^{-}$¨ion indicate clearly that our variational expansion Eq.(\ref{equa1}) is very effective in 
applications to four-electron atomic systems. In this Section we apply the same variational expansion, Eq.(\ref{equa1}), to investigate another long-standing problem known in the atomic physics 
of four-electron atomic systems. Briefly, our goal is to explain variations of the half-life of the beryllium-7 isotope in different chemical enviroments. As follows from the results of 
numerous experiments, the half-life of the ${}^{7}$Be isotope is `chemically dependent', i.e. it varies by $\approx$ 0.5 \% - 5 \% for different chemical compounds. This fact contradicts an old 
fundamental statement (see, e.g., \cite{Remi}) that actual decay rates of chemical isotopes cannot depend upon their chemical enviroments. This explains a substantial interest in chemical 
compounds which contain atoms of the beryllium-7 (or ${}^{7}$Be) isotope. It should be mentioned that in modern laboratories different chemical compounds containing ${}^{7}$Be atoms 
are not `exotic' substances, since the nuclei of ${}^{7}$Be are formed in the $(p;n)-$ and $(p;\alpha)-$reactions of the ${}^{7}$Li and ${}^{10}$B nuclei with the accelerated protons. A few other 
nuclear reactions involving nuclei of some light and intermediate elements, e.g., C, Al, Cu, Au, etc, also lead to the formation of ${}^{7}$Be nuclei. In general, an isolated ${}^{7}$Be nucleus 
decays by using a few different channels, the most important of which is the electron capture (or $e^{-}-$capture) of one atomic electron from the internal $1s^2-$shell. The process is described 
by a simple atomic-nuclear equation ${}^{7}$Be $\rightarrow$ ${}^{7}$Li, where there is no free electron emitted after the process. During this process the maternal ${}^{7}$Be nucleus is 
transformed into the ${}^{7}$Li nucleus which can be found either in the ground state, or in the first excited state. The subsequent transition of the excited ${}^{7}$Li$^{*}$ nucleus into its 
ground state ${}^{7}$Li proceeds with the emission of a $\gamma-$quantum which has energy $E_{\gamma} \approx 0.477$ $MeV$. Such $\gamma-$quanta can easily be registered in modern experiments 
and this explains numerous applications of chemical compounds of ${}^{7}$Be in radio-chemistry.      

Let us discuss the process of the electron capture in the ${}^{7}$Be-atom in detail. Assume for a moment that all ${}^{7}$Be atoms decay by electron capture from the ground (atomic) 
$2^{1}S-$state. In this case, by using the expectation value of the electron-nucleus delta-function $\langle \delta({\bf r}_{eN}) \rangle$ computed for the ground $2^1{}S-$state of an 
isolated Be-atom we can write the following expression for the half-life $\tau$ of the ${}^{7}$Be atom/isotope
\begin{equation}
   \tau = \frac{1}{\Gamma} = \frac{1}{A \langle \delta({\bf r}_{eN}) \rangle} \; \; \; \label{eq1}
\end{equation}
where $\Gamma$ is the corresponding width and $A$ is an additional factor which in principle depends on the given chemical compound of beryllium. The half-life $\tau$ determines the moment when 
50 \% of the incident ${}^{7}$Be will have decayed by electron capture. An analytical formula for $\tau$, Eq.(\ref{eq1}), follows from the fact that the corresponding width $\Gamma = \tau^{-1}$ 
must be proportional to the product of theexpectation value of the electron-nucleus delta-function and an additional factor $A$. The expectation value of the electron-nucleus delta-function 
computed with the non-relativistic wave function determines the electron density at the surface of a sphere with the spatial radius $R \approx \Lambda_e = \frac{\hbar}{m_e c} a_0 = \alpha a_0$, 
where $a_0$ is the Bohr radius $a_0 \approx \frac{\hbar^2}{m_e e^2} (\approx 5.292 \cdot 10^{-9}$ $cm$), $c$ is the speed of light and $\Lambda_e$ is the Compton wave length. The `constant' $A$ 
in Eq.(\ref{eq1}) represents an `additional' probability for an electron (point particle) to penetrate from the distance $R \approx \Lambda_e = \alpha a_0$ to the surface of the nucleus $R_N 
\approx 1 \cdot 10^{-13}$ $cm$. 

A numerical value of $A$ can be evaluated by assuming that the mean half-life of the ${}^{7}$Be-atom in its ground $2^{1}S-$state equals 53.60 days and by using our best expectation value obtained 
for the expectation value of the electron-nucleus delta-function $\langle \delta({\bf r}_{eN}) \rangle \approx$ 8.82515 $a.u.$, one finds that $\Gamma \approx 2.1593422 \cdot 10^{-7}$ sec$^{-1}$. 
From here we find that the factor $A$ in Eq.(\ref{eq1}) equals
\begin{equation}
  A \approx \frac{2.1593422 \cdot 10^{-7}}{\langle \delta({\bf r}_{eN}) \rangle} \approx 2.439521 \cdot 10^{-8} \; \; \;  \label{eq2}
\end{equation}
where the expectation value $\langle \delta({\bf r}_{eN}) \rangle$ must be taken in atomic units. As follows from numerous experiments the mean life-timed of chemical compounds which contain some 
${}^{7}$Be-atom(s) are $\approx$ 53 - 54 days. This means that the `constant' $A$ varies slowly in actual molecules. This allows us to write the following approximate formula for the ratio of 
half-life of the two different molecules X(Be) and Y(Be) which contain ${}^{7}$Be atoms
\begin{equation}
   \frac{\tau({\rm X(Be)})}{\tau({\rm Y(Be)})} = \frac{\langle \delta({\bf r}_{eN}); {\rm Y(Be)} \rangle}{\langle \delta({\bf r}_{eN}); {\rm X(Be)} \rangle} 
   \; \; \;  \label{eq3}
\end{equation}
Let us apply this formula to the case when one of the ${}^7$Be-atoms is in the ground $2^1S-$state, while another such an atom is in the triplet $2^{3}S-$state. The expectation value of the 
$\delta_{eN}$-function for the ground state in the Be-atom is given above, while for the triplet state we have $\langle \delta({\bf r}_{eN}) \rangle \approx$ 8.739558 $a.u.$ Both these expectation 
values were determined in our highly accurate computations of the ground $2^{1}S-$ and $2^{3}S-$state in the four-electron Be atom. With these numerical values one finds from Eq.(\ref{eq3}) that 
the half-life of the ${}^{7}$Be atom in its triplet $2^3S-$state is 1.009794 times (or by $\approx$ 1 \%) longer than the corresponding half-life of the ${}^{7}$Be atom in its ground singlet 
$2^1S$-state. This simple example includes two different bound states in an isolated ${}^{7}$Be-atom. In general, by using the formula Eq.(\ref{eq3}) we can approximately evaluate the half-life of 
the ${}^{7}$Be atoms in different molecules and compounds. The formula Eq.(\ref{eq3}) can be applied, e.g., to BeO, BeC$_2$, BeH$_2$ and many other beryllium compounds, including beryllium-hydrogen 
polymers, e.g., Be$_n$H$_{2n}$ for $n \approx 100 - 1000$ (see, e.g., \cite{Ref4} - \cite{Ref7} and references therein).

As is well known from atomic physics, the electronic structure of the excited bound states of the four-electron  Be-atom(s) is $1s^2 2s n\ell$ (or $1s^2 2s^1 n\ell^1$), where $\ell \ge 0$ and $n \ge 3$. 
In general, such an excited state arises after excitation of a single electron from the $1s^2 2s^2$ electron configuration, which correspond to the ground state, or `core', for short. It is clear that 
the final $1s^2 2s^1 n\ell^1$ configuration is the result of a single electron excitation $2s \rightarrow n \ell$. All other states with excitation(s) of two and more electrons from the core are unbound. 
In general, a very substantial contribution ($\ge$ 95 \%) to the expectation value of the electron-nucleus delta-function comes from the two internal electrons (or $1s^2-$electrons) of the Be-atom.  
Briefly this means that the expectation value of the electron-nucleus delta-function is almost the same for all molecules which contain the bound Be-atom. Variations in 3 \% - 6 \% are possible and they
are related with the contribution of the  two outer-most electrons in the expectation value of the electron-nucleus delta-function $\langle \delta({\bf r}_{eN}) \rangle$. As follows from computational 
results the overall contribution from two outer-most electrons is only 3 \% - 6 \% of the total numerical value. This means that variations in the chemical enviroment of one ${}^{7}$Be atom can change the 
half-life of this atom by a factor of 1.03 to 1.06 (maximum). In reality, such changes are significantly smaller, but they can be noticed in modern experiments. 

It is interesting to note that analogous result (3 \% - 6 \% differences as maximum) can be predicted for other nuclear processes which are influenced by variations in the chemical enviroment, e.g., for 
the excitation of the ${}^{235}$U nucleus which also depends upon chemical  enviroment \cite{XX} - \cite{Fro2005}. It is well known (see, e.g., \cite{ZZ}) that the ${}^{235}$U nucleus 
has an excited state with the energy $\approx$ 75 - 77 $eV$. There is no such level in the ${}^{234}$U, ${}^{236}$U and ${}^{238}$U nuclei. Nuclear properties of the ground and first excited states in 
the ${}^{235}$U nucleus differ substantially. Moreover, by changing the actual chemical enviroment of the ${}^{235}$U atom we can change the probabilities of excitation of the central nucleus, e.g., by 
using different alloys of uranium, in order to change and even control nuclear properties. For instance, this approach can be used to achieve and even exceed critical conditions with respect to neutron 
fission. Theoretical evaluations and preliminary experiments show that possible changes in nuclear properties of different compounds of uranium-235 do not exceed 3 - 6 \%. It is very likely that 3 - 6 \% 
is the upper limit of influence of atomic (and molecular) properties on the nuclear properties of different isotopes. On the other hand, possible changes in atomic and molecular properties produced by 
processes, reactions and decays in atomic nuclei are always significant.    

Thus, if we know the expectation value of the electron-nucleus delta-function for the beryllium-7 atom within some molecule with other chemical elements, then we can evaluate the corresponding half-life 
of such an atom with respect to electron capture. Currently, however, this problem can be solved only approximately, since there are quite a few difficulties in accurate computations of complex 
molecules as well as in actual experiments, since, e.g., the exact value of the constant $A$ in Eq.(\ref{eq1}) is not known. In other words, we cannot be sure that the experimental half-life mentioned 
above (53.60 days) corresponds to the electron capture in the ground $2^{1}S-$state of an isolated ${}^{7}$Be atom. In fact, it is not clear what chemical compounds were used (and at what conditions) 
to obtain this half-life. Very likely, we are dealing with some `averaged' value determined for a mixture of different molecules. It is clear that improving the overall experimental accuracy and purity 
of future experiments is critical. The accuracy of future theoretical computations could also be improved. First of all, we need to focus on accurate expectation values of the electron-nucleus 
delta-function $\langle \delta({\bf r}_{eN}) \rangle$, rather than just accurate values of the total energy. Then the formula, Eq.(\ref{eq3}), can be used to determine the actual life-times of the 
${}^{7}$Be atoms, which are included in different chemical compounds.       

\section{Conclusion}

We have considered the bound state properties of the negatively charged Li$^{-}$ ion in the ground $2^1S-$state. These bound state properties are compared with the analogous properties of the neutral 
Li atom. Our analysis of the bound state properties of the Li$^{-}$ ion is of interest since the formation of the negatively charged Li$^{-}$ ions plays an important role in modern lithium and 
lithium-ion batteries. Expectation values of different properties determined in this study are sufficient for all current and anticipated future experimental needs. As follows from the results of our 
calculations the Li$^{-}$ ion is a weakly-bound atomic system which has only one bound $2^{2}S-$state. The internal structure of this state is represented as a motion of one `almost free' electron in 
the field of a heavy atomic cluster which is the neutral Li atom in its ground $2^2S-$state. The computed expectation values of the bound state properties of the Li$^{-}$ ion in the ground 
$2^1S-$state and the neutral Li atom in the ground $2^2S-$state support such a picture. Moreover, the whole internal structure of the Li$^{-}$ ion could be reconstructed to very good accuracy if we 
knew the model potential between an electron and neutral Li atom. This corresponds to the two-body approximation which is often used for weakly bound few-body systems. An accurate reconstruction of 
such a model $e^{-}$-Li interaction potential should be a goal of future research. The same model potential could then be used to obtain the cross-section of the elastic scattering (at relatively 
small energies) for the electron-lithium scattering.  

It should be mentioned that the negatively charged ${}^{6}$Li$^{-}$ ion is of interest for possible creation and observation of an unstable (three-electron) ${}^{4}$He$^{-}$ ion which is formed in 
one of the channels of the reaction of the ${}^{6}$Li$^{-}$ ion with slow neutrons, e.g., 
\begin{eqnarray}
 {}^6{\rm Li}^{-} + n = {}^4{\rm He}^{-} + {}^3{\rm H}^{+} + e^{-} + 4.785 \; MeV \; \; \; , \label{concl1}
\end{eqnarray}
Preliminary evaluations indicate that the probability of formation of the ${}^{4}$He$^{-}$ ion in this reaction is $\approx$ 0.02 \% - 0.04 \%. Nevertheless, this nuclear reaction of the 
${}^6$Li$^{-}$ ion with slow neutrons has a very large cross-section and it can be used to produce the negatively charged He$^{-}$ ion which is unstable and decays into the neutral He atom with the 
emission of one electron. Other approaches to create relatively large numbers of the negatively charged ${}^{4}$He$^{-}$ ions have failed. 

We also investigated the situation of experimental variations of the half-life of the beryllium-7 isotope placed in different chemical enviroments. Since the middle of the 1930's this interesting 
problem has attracted significant experimental and theoretical attention. It is shown here that the half-life of the beryllium-7 isotope in different chemical enviroments may vary by 3 \% - 6 \% 
(maximum). A central computational part of this problem is to determine to high accuracy the electron-nucleus delta-function of the Be-atom placed in different molecules, `quasi-metalic' alloys and 
other chemical compounds. The currently achieved accuracy is not sufficient to make accurate predictions of the half-life of the beryllium-7 atom in many molecules. Another part of the solution is 
to improve the accuracy and the purity of the chemical enviroment in all experiments performed with different molecules which include atoms of beryllium-7.  

\begin{center}
  {\Large Appendix} \\
\end{center}

The expectation values $\langle {\bf p}_e \cdot {\bf p}_N \rangle$ and $\langle {\bf p}_e \cdot {\bf p}_e \rangle$ are not presented in Table I, since they are not truly independent from the $\langle 
\frac12 p^2_e \rangle$ and $\langle \frac12 p^2_N \rangle$ expectation values which are given in Table I. Indeed, for an arbitrary $K-$electron atom/ion the expectation values of the scalar products 
of the vectors of electron's momenta ${\bf p}_i$ ($i = 1, \ldots, K$) with the  electron's momenta ${\bf p}_i$ ($j \ne i$ and $j = 1, \ldots, K$) and with the momentum of the nucleus ${\bf p}_N$ are 
simply related with the expectation values of the single-electron kinetic energy and kinetic energy of the atomic nucleus:
\begin{eqnarray}
 & & \langle {\bf p}_i \cdot {\bf p}_j \rangle = \langle {\bf p}_1 \cdot {\bf p}_2 \rangle = \frac{2}{K (K - 1)} \Bigl[ \langle \frac12 p^2_N \rangle - 2 \langle \frac12 p^2_e \rangle \Bigr] 
 \; \; \; \label{App1} \\
 & & \langle {\bf p}_i \cdot {\bf p}_N \rangle = \langle {\bf p}_1 \cdot {\bf p}_N \rangle = - \frac{2}{K} \langle \frac12 p^2_N \rangle \; \; \; \; , \; \label{App2} 
\end{eqnarray}
where $K (\ge 2)$ is the total number of electrons in atom, $\langle {\bf p}_i \cdot {\bf p}_j \rangle$ is the scalar product of the two electron momenta ($i \ne j$), while $\langle {\bf p}_i \cdot 
{\bf p}_N \rangle$ is the scalar product of the momenta of the atomic nucleus and electron (with index $i$). Since the electron's indexes can be chosen arbitrarily, we can replace the scalar products in 
Eqs.(\ref{App1}) - (\ref{App2}) by the $\langle {\bf p}_1 \cdot  {\bf p}_2 \rangle$ and $\langle {\bf p}_1 \cdot  {\bf p}_N \rangle$ expectation values, respectively. In general, these two expectation 
values determine the electron-electron and electron-nucleus kinematic correlations in few- and many-electron atoms. In Eqs.(\ref{App1}) and (\ref{App2}) the notations $\langle \frac12 p^2_e \rangle$ 
and $\langle \frac12 p^2_N \rangle$ designate the single-electron kinetic energy and kinetic energy of the atomic nucleus, respectively. Therefore, there is no need to include the $\langle {\bf p}_1 
\cdot {\bf p}_2 \rangle$ and $\langle {\bf p}_1 \cdot {\bf p}_N \rangle$ expectation values in Table I. Also, it is interesting to note that the nuclear charge $Q$ is not included in Eqs.(\ref{App1}) 
- (\ref{App2}). This means that Eqs.(\ref{App1}) - (\ref{App2}) can be applied to an arbitrary $K-$electron atom, or positevely/negatively charged ion. For two-electron atomic systems we have $K = 2$ 
and Eqs.(\ref{App1}) - (\ref{App2}) mentioned above take the well known form (see, e.g., \cite{Eps}, \cite{Fro2007})
\begin{equation}
 \langle {\bf p}_1 \cdot {\bf p}_2 \rangle = \langle \frac12 p^2_N \rangle - 2 \langle \frac12 p^2_e \rangle \; \; \; , \; \; \; \langle {\bf p}_e \cdot {\bf p}_N \rangle = \langle {\bf p}_1 
 \cdot {\bf p}_N \rangle = - \langle \frac12 p^2_N \rangle \; \label{App3} 
\end{equation}

\newpage


 \begin{table}[tbp]
   \caption{The expectation values of a number of electron-nuclear ($en$) and electron-electron ($ee$) properties (in $a.u.$) of the ground $2^{1}S-$ and $2^2S-$states 
            of the of the Li$^{-}$ (${}^{\infty}$Li$^{-}$) ion and neutral Li (${}^{\infty}$Li) atom, respectively.}
     \begin{center}
     \begin{tabular}{| c | c | c | c | c | c | c | c |}
       \hline\hline          
 atom/ion  & state  & $\langle r^{-2}_{eN} \rangle$ & $\langle r^{-1}_{eN} \rangle$ & $\langle r_{eN} \rangle$ & $\langle r^2_{eN} \rangle$  & $\langle r^3_{eN} \rangle$ & $\langle r^4_{eN} \rangle$ \\
     \hline
 Li$^{-}$        & $2^1S$ & 7.56810    & 1.47465   & 2.90556   & 17.539   & 140.8   & 1355 \\

 Li              & $2^2S$ & 10.0803050 & 1.9060373 & 1.6631655 & 6.118051 & 30.8650 & 183.318 \\
       \hline\hline     
 atom/ion  & state & $\langle r^{-2}_{ee} \rangle$ & $\langle r^{-1}_{ee} \rangle$ & $\langle r_{ee} \rangle$ & $\langle r^2_{ee} \rangle$ &  $\langle r^3_{ee} \rangle$ &  $\langle r^4_{ee} \rangle$ \\
     \hline
 Li$^{-}$        & $2^1S$ & 0.74750  & 0.44883 & 5.1253  & 38.127  & 352.81 & 3832 \\

 Li              & $2^2S$ & 1.46039553 & 0.7327379 & 2.8894478 & 12.28230 & 64.0240 & 385.173 \\
      \hline\hline             
 atom/ion  & state & $E$ & $\langle \frac12 p^2_{e} \rangle$ & $\langle \frac12 p^2_{N} \rangle$ & $\langle \delta_{eN} \rangle$ & $\langle \delta_{ee} \rangle$ &  $\langle \delta_{eee} \rangle$  \\ 
     \hline
 Li$^{-}$        & $2^1S$ & -7.5007605  & 1.875509   & 7.809830   & 3.42829   & 9.1421$\times 10^{-2}$   & 0.0 \\

 Li              & $2^2S$ & -7.47800737 & 2.49268725 & 7.77990315 & 4.607933  & 0.181640   & 0.0 \\
     \hline\hline
  \end{tabular}
  \end{center}
   \end{table}

 \begin{table}[tbp]
   \caption{The total energies and some electron-nuclear ($eN$) and electron-electron ($ee$) properties in $a.u.$ of 
            a few selected heavy three-electron ions in their the ground $2^2S-$state(s).} 
     \begin{center}
     \begin{tabular}{| c | c | c | c | c |}
       \hline\hline          
  ion                          & Sc$^{18+}$      & Ti$^{19+}$      & V$^{20+}$       & Cr$^{21+}$  \\
          \hline
                          $E$  & -475.0551425155 & -522.4072925498 & -572.0094468708 & -623.8616048933 \\ 
                            \hline
 $\langle r^{-2}_{eN} \rangle$ &  604.22639 &  664.15054 &  726.90919 &  792.50086 \\ 

 $\langle r^{-2}_{ee} \rangle$ &  110.11954 &  121.22245 &  132.85859 &  145.02796 \\
                   \hline
 $\langle r^{-1}_{eN} \rangle$ &  15.409047 &  16.159048 &  16.909049 &  17.659050 \\

 $\langle r^{-1}_{ee} \rangle$ &  6.8866056 &  7.2275887 &  7.5685671 &  7.9095412 \\
                 \hline
 $\langle r_{eN} \rangle$      & 0.14981991 & 0.14268792 & 0.13620461 & 0.13028523 \\

 $\langle r_{ee} \rangle$      & 0.24566863 & 0.23390439 & 0.22321615 & 0.21346256 \\
                 \hline\hline
  ion                          &    Mn$^{22+}$   &    Fe$^{23+}$   &    Co$^{24+}$   &     Ni$^{25+}$  \\
          \hline
                         $E$   & -677.9637661344 & -734.3159301916 & -792.9180967274 & -853.7702654564 \\
                           \hline
 $\langle r^{-2}_{eN} \rangle$ &  860.92690 &  932.18472  &  1006.2789 &  1083.2040 \\

 $\langle r^{-2}_{ee} \rangle$ &  157.73072 &  170.96695  &  184.73616 &  199.03891 \\
                           \hline 
 $\langle r^{-1}_{eN} \rangle$ &  18.409051 &  19.159052  &  19.909053 &  20.659053 \\

 $\langle r^{-1}_{ee} \rangle$ &  8.2505119 &  8.5914772  &  8.9324449 &  9.2734076 \\
                              \hline
 $\langle r_{eN} \rangle$      & 0.12485925 & 0.11986738  & 0.11525955 & 0.11099301 \\

 $\langle r_{ee} \rangle$      & 0.20452616 & 0.19630825  & 0.18872561 & 0.18170715 \\
       \hline\hline     
  \end{tabular}
  \end{center}
   \end{table}
\begin{table}[tbp]
   \caption{The total non-relativistic energies $E$ of the different atoms/ions in 
            their ground (bound) states in atomic units. All nuclear masses are infinite. 
            $Q$ is the electric charge of the atomic nucleus and $N_e$ is the total number 
            of bounded electrons. All energies for the He-like atoms/ions and some energies 
            of the Li-like ions (after $Q$ = 20) have been determined in this study. This 
            Table is useful for accurate eveluations of binding energies, relativistic 
            corrections, etc in few-electron atoms/ions.}
     \begin{center}
     \scalebox{0.80}{%
     \begin{tabular}{| c | c | c | c |}
      \hline\hline
  $Q$ &    He-like $(N_e = 2)$               & Li-like ($N_e = 3$) & Be-like ($N_e = 4$)    \\
     \hline\hline
   1 &   -0.5277510165443771965925           &  -----------------  &  ---------------- \\      
   2 &   -2.90372437703411959831115924519440 &  -----------------  &  ---------------- \\  
   3 &   -7.27991341266930596491875          &   -7.4780603236503  &  ---------------- \\  
   4 &  -13.65556623842358670208051          &  -14.3247631764654  &  -14.667356407951 \\
   5 &  -22.03097158024278154165469          &  -23.424605720957   &  -24.348884381902 \\
   6 &  -32.40624660189853031055685          &  -34.775511275626   &  -36.534852285202 \\
             \hline 
   7 &  -44.781445148772704645183         &  -48.376898319137   &  -51.222712616143 \\
   8 &  -59.156595122757925558542         &  -64.228542082701   &  -68.411541657589 \\
   9 &  -75.531712363959491104856         &  -82.330338097298   &  -88.100927676354 \\
  10 &  -93.906806515037549421417         & -102.682231482398   & -110.290661070069 \\
  11 & -114.28188377607272189582          & -125.2841907536473  & -134.980624604257 \\
  12 & -136.65694831264692990427          & -150.1361966044594  & -162.170747906692 \\
             \hline 
  13 & -161.03200302605835987252          & -177.238236559961   & -191.860986338262 \\
  14 & -187.40704999866292631487          & -206.5903022122780  & -224.051310298012 \\
  15 & -215.78209076353716023462          & -238.1923876941461  & -258.741699427160 \\
  16 & -246.15712647425473932009          & -272.0444887900725  & -295.932139288646 \\
  17 & -278.53215801540009570337          & -308.1466023952556  & -335.622619375075 \\
  18 & -312.90718607661114879880          & -346.4987261736714  & -377.813131866050 \\
             \hline 
  19 & -349.28221120345316700447          & -387.1008583345610  & -422.503670826658 \\
  20 & -387.65723383315855621790          & -429.9529974827626  & -469.694231675265 \\
  21 & -428.03225432023469116264          & -475.0551425155     & -519.384810821074 \\
  22 & -470.40727295513838395930          & -522.4072925498     & -571.575405411671 \\
  23 & -514.78228997811177388135          & -572.0094468708     & -626.266013153662 \\
  24 & -561.15730558958127234352          & -623.8616048933     & -683.456632182920 \\
  25 & -609.53231995807574620568          & -677.9637661344     & -743.147260969064 \\
              \hline 
  26 & -659.90733322632780520901          & -734.3159301916     & -805.337898245040 \\
  27 & -712.28234551602655145614          & -792.9180967274     & -870.028542951686 \\
  28 & -766.65735693155709991040          & -853.7702654564     & -937.219194199135 \\
  30 & ------------------------           &  -----------------  & -1079.100513407098 \\
  36 & ------------------------           &  -----------------  & -1564.744568198454 \\
     \hline\hline
  \end{tabular}}
  \end{center}
  \end{table}
\end{document}